\documentclass[conference]{IEEEtran}
\usepackage[linesnumbered,ruled,vlined]{algorithm2e}
\usepackage{amsmath}
\usepackage{epsfig,amssymb,subfigure,bm,dsfont}
\usepackage{hyperref}
\usepackage{algpseudocode}
\usepackage{amssymb}
\usepackage{amsfonts}
\usepackage{array}
\usepackage{balance}
\usepackage{bbding}
\usepackage{booktabs}
\usepackage{calc}
\usepackage{caption}
\usepackage{cite}
\usepackage{color}
\usepackage{curves}
\usepackage{epsfig}
\usepackage{epstopdf}
\usepackage{epic}
\usepackage{float}
\usepackage{footnote}
\usepackage{graphicx}
\usepackage{mathrsfs}
\usepackage{mdwlist}
\usepackage{multicol}
\usepackage{ntheorem}
\usepackage{pifont}
\usepackage{psfrag}
\usepackage{url}
\usepackage{subfigure}
\usepackage{stfloats}
\usepackage{stmaryrd}
\usepackage{times}
\usepackage{tipa}
\usepackage[english]{babel}
\begin{document}
\bstctlcite{IEEEexample:BSTcontrol}
\theoremheaderfont{\bfseries\upshape}
\theoremseparator{:}
\newtheorem{proof}{Proof}
\newtheorem{theorem}{Theorem}
\newtheorem{assumption}{Assumption}
\newtheorem{proposition}{Proposition}
\newtheorem{definition}{Definition}
\newtheorem{lemma}{Lemma}
\newtheorem{corollary}{Corollary}
\newtheorem{remark}{Remark}
\newtheorem{construction}{Construction}
\newtheorem{problem}{Problem}
\newtheorem{alg}{Algorithm}[section]
\captionsetup{font={scriptsize}}
\renewcommand{\IEEEQED}{\IEEEQEDclosed}

\newcommand{\supp}{\mathop{\rm supp}}
\newcommand{\sinc}{\mathop{\rm sinc}}
\newcommand{\spann}{\mathop{\rm span}}
\newcommand{\essinf}{\mathop{\rm ess\,inf}}
\newcommand{\esssup}{\mathop{\rm ess\,sup}}
\newcommand{\Lip}{\rm Lip}
\newcommand{\sign}{\mathop{\rm sign}}
\newcommand{\osc}{\mathop{\rm osc}}
\newcommand{\R}{{\mathbb{R}}}
\newcommand{\Z}{{\mathbb{Z}}}
\newcommand{\C}{{\mathbb{C}}}
\newcommand*{\affaddr}[1]{#1} % No op here. Customize it for different styles.
\newcommand*{\affmark}[1][*]{\textsuperscript{#1}}
\newcommand*{\email}[1]{\texttt{#1}}

%=======================================================title==============================================================================
\title{Global Differential Privacy for Distributed Metaverse Healthcare Systems}
%\author{{Mehdi Letafati$^\ast$, Moayad Aloqaily$^\dagger$, and Safa Otoum$^\ast$ \\ \vspace{3mm}  
%\small{$^\ast$Zayed University, Dubai, UAE\\  } 
%\small{$^\dagger$Mohamed Bin Zayed University of Artificial Intelligence, Abu Dhabi, UAE}
%}
%}

	\author{%
		\IEEEauthorblockN{%
Mehdi Letafati\IEEEauthorrefmark{1} and 
%Moayad Aloqaily\IEEEauthorrefmark{2}, 
Safa Otoum\IEEEauthorrefmark{1}		          
}% 
	%\\
\IEEEauthorblockA{\IEEEauthorrefmark{1}   Zayed University, Dubai, UAE}%
	%	\IEEEauthorblockA{\IEEEauthorrefmark{1} \small Computer Engineering Department, Sharif University of Technology, Tehran, Iran}%
%\IEEEauthorblockA{\IEEEauthorrefmark{2}  Mohamed Bin Zayed University of Artificial Intelligence (MBZUAI), Abu Dhabi, UAE}%
\IEEEauthorblockA{\small Emails:‌ $^\ast$mletafati@ee.sharif.edu; $^\ast$safa.otoum@zu.ac.ae;   %$^\dagger$moayad.aloqaily@mbzuai.ac.ae
}
}

\maketitle 

\IEEEaftertitletext{\vspace{-2.5\baselineskip}}

%==============================================System model=====================================================================
\begin{abstract}
Metaverse-enabled  digital healthcare systems are expected to exploit  an unprecedented amount of personal health data, while ensuring that  sensitive or private information of individuals are not disclosed. Machine learning and artificial intelligence (ML/AI) techniques can be widely utilized in metaverse healthcare systems, such as virtual clinics and intelligent consultations.  
In such scenarios, the key challenge is that data privacy laws might not allow virtual clinics to share their medical data with other parties. Moreover, clinical AI/ML models themselves carry extensive information about the medical datasets, such that private attributes can be easily inferred by malicious actors in the metaverse (if not rigorously privatized).   
In this paper, inspired by the idea of ``incognito mode'', which has recently been developed as a promising solution to safeguard  metaverse users' privacy,  
%based on client-level  $\epsilon$-differential privacy, in this paper  
we propose  \emph{global differential privacy} for the  distributed metaverse healthcare systems.    
%Each intelligent healthcare  party perturbs its model parameters,  and  sends the randomized version to the aggregator node, to  enhance the privacy against both the malicious actors  and curious servers.
In our scheme, a randomized mechanism in the format of artificial ``mix-up'' noise is applied to the federated clinical ML/AI models 
%(occupied  by each of the virtual medical centers in the metaverse) 
before sharing with other peers.  This way, we provide an adjustable level of \emph{distributed privacy} against both the malicious actors and honest-but-curious metaverse servers.   
Our evaluations on breast cancer Wisconsin dataset (BCWD) highlight the privacy-utility trade-off (PUT) in terms of diagnosis accuracy and loss function for different levels of privacy.  We also  compare our private scheme with the  non-private  centralized setup in terms of diagnosis accuracy.  
\end{abstract}

\begin{IEEEkeywords}
Distributed differential privacy, globally-private metaverse, digital healthcare, distributed learning, federated learning.   
\end{IEEEkeywords}

\section{Introduction} 
Metaverse can be considered as the convergence point of technologies that facilitate immersive  interactions between  physical world and  virtual environments \cite{Metav1, Metav2, Moayad_BC}. 
Digital healthcare services, as one of the major  applications of the metaverse,  aims  to provide seamless interactions between the patients and physicians for medical diagnosis and treatment, therapy sessions, etc \cite{Jamshid, Moayad_DT, Moayad_BC}.  

\subsection{Motivation \& Background}
In a metaverse healthcare  platform, patients and physicians use wearables or  augmented/virtual reality  (AR/VR) devices with  built-in sensors in order to render a high quality immersive experience for the healthcare service.  For instance, Surgeons may utilize AR for surgery by augmenting the part of body where it needs the procedure, while collected data from  the real world  are exploited for visualization  in the virtual space.   These devices act as a ``gateway'' to the metaverse era,  collecting  bio-metric data, brain patterns, users' speech, facial expressions, body movements and poses, etc. 
%Such a miscellaneous set of sensitive data and information flow in the context of metaverse  provides adversaries with a broad attacking surface which needs to be carefully addressed by considering  novel secrecy- and privacy-preserving mechanisms 
%to be employed by  network designers  who aim to offer e-health platforms for the metaverse 
%\cite{WSKG-letter, WSKG-GC, vtc2022, arxive, ICC}.   
This can create new venues for both passively and actively attacking the access layer to the metaverse which requires scalable  solutions for securing the access with  new principles of cyber-resilience,  while bringing continuous secrecy update \cite{WSKG-letter, WSKG-GC, vtc2022, arxive, ICC}.   
In addition to that,  machine learning and artificial intelligence (ML/AI) are widely  utilized in healthcare to help facilitate accurate medical diagnosis and treatment, drug development, etc., in virtual clinics and intelligent  metaverse-enabled consultations \cite{battle_privacy}.   In this case, the point is that data privacy laws may not allow virtual
clinics or medical centres in the metaverse to share their medical data with
others. Besides,  model updates in clinical AI/ML algorithms contain  information about the medical datasets, and malicious participants in the metaverse can 
sniff  the packets containing  model parameters to  infer private attributes  \cite{model_inversion,  medical_deepfake}.  
In terms of the employed ML/AI
algorithms in the metaverse, most healthcare systems are based
on centralized approaches, which, in general, can pose several
potential threats such as security and robustness flaws (e.g.,
single point of failure) and privacy leakage to data aggregator
entities.  

%In the metaverse healthcare systems, highly accurate predictive healthcare systems can be developed by exploiting AI/ML technologies. Physicians and medical doctors can leverage AI algorithms (fed with rich medical datasets) to
%model and train for challenging surgeries attempting on a
%real case. To mention other types of ML/AI-powered health-
%related services in the metaverse, computer vision-aided action
%and gesture recognition systems can be considered, since
%avatars need to understand the actions of other avatars in
%the virtual environment. 

Although  metaverse can bring amazing services to patients,  health professionals, and medical clinics,  the employed AI/ML algorithms are highly  vulnerable to  privacy and security risks.  
The big concern is that the learning algorithms might leak users' personal data \cite{model_inversion, backdoor}. In addition,  avatars and virtual clinics or medical centres might not be willing to share the health data in the metaverses. This  makes it challenging for AI/ML algorithms to carry out a comprehensive data analysis for enhancing the healthcare  services in  the metaverse.      
In this context, digital healthcare  service providers should seek for \emph{distributed learning} approaches for the metaverse.  
As such, the generic  framework of federated learning (FL) has shown to be a promising ``privacy-aware'' (not necessarily privacy-preserving)  solution for running distributed learning protocols \cite{FL_Quek}. Specifically, 
 when training the medical ML/AI models with large-sized  data and images, e.g.,  high-resolution video streams or 3D images, FL can significantly reduce the    overhead of collaborative learning, since the medical model parameters are simply  exchanged with the coordinator entities, but not the private raw data. 
 %Moreover,  for e-health use-case as a data-sensitive application, data privacy laws may not allow clinics or e-health intelligent agents to share their medical data  with others.  Therefore,  FL protocols  can be a promising candidate for deployment in this case.  

To address the privacy risks of AI/ML algorithms, the integration of  FL and cross-chain technologies is considered in \cite{BC_FL_core} as a blockchain-based framework for distributed learning in the metaverse.  
 A main chain is employed to render the parameter server (PS), while  multiple subchains are implemented to manage local model updates.  
 %generated by smart devices (or their digital counterpart) acting as workers.   
Although a considerable number of papers address the blockchain technology as an inherently secure 
	%as an inherent means of realizing  privacy- and secrecy-aware platforms  
	 framework  to  realize privacy in the metaverse (See \cite{Moayad_BC} and references therein),   recent studies  clearly  demonstrate  serious vulnerabilities in terms of  privacy in the current metaverse platforms \cite{Vivek1, Vivek2, Vivek3}. 
%  demonstrate  that more than 50000 can be uniquely
%and reliably identified across multiple sessions using just their head and hand motion relative to virtual objects. 
It is shown in \cite{Vivek3}  that after training a centralized classification model  using only  5 minutes of data per person,  each metaverse  user can be uniquely identified amongst the entire pool of 50,000+  with 73.20\% accuracy from just 10 seconds of being active in the metaverse,  and with 94.33\% accuracy from 100 seconds of motion.  
As another study \cite{Vivek1},  over 25 personal data attributes  of fifty participants were inferred 
%playtested an innocent-looking “escape room” game in virtual reality (VR).
within just a few minutes of participating in the metaverse games and fitness-related  activities.  
To deal with the unprecedented  privacy risks in the metaverse, the idea of “incognito mode” has recently been proposed in \cite{Vivek2} as a novel solution  for safeguarding the privacy of VR users in the metaverse.   
To realize the incognito mode in the metaverse,  client-level local $\epsilon$-differential privacy (DP) framework is proposed by the researchers 
%which has been  shown to be capable of 
to protect  a wide variety  of sensitive data attributes for VR applications.  Local DP (with adjustable privacy level $\epsilon$) applies randomized responses, noise, or data perturbations (depending on the format of the data attributes we aim to privatize) on the client side, in contrast to on the server side.

 \subsection{Contributions \& State-of-the-Arts}
{In this paper, we are inspired by the work of \cite{Vivek2}, where we  generalize the concept of  ``client-level'' differentially-private  metaverse, and  propose  \emph{global differential privacy} for distributed learning-based  metaverse healthcare systems.    
%Each intelligent healthcare  party perturbs its model parameters,  and  sends the randomized version to the aggregator node, to  enhance the privacy against both the malicious actors  and curious servers.
In our scheme, randomized mechanism is realized in  the form of ``mix-up'' noise which is applied to the federated clinical ML/AI models---each of the virtual medical centers  perturbs its clinical model  parameters by adding mix-up noises before sharing them with the parameter server for aggregation. Then in the aggregation phase, this perturbation is adaptively fine-tuned to meet the privacy requirements  during global model update. Our global  $(\epsilon, \delta)$-DP framework can provide an adjustable level of  privacy against both the malicious actors and honest-but-curious metaverse servers, i.e., the servers which follow the network protocols,  while at the same time  might try to obtain some  information  about users from the  data they receive.  
We evaluate our scheme on breast cancer Wisconsin dataset (BCWD)\footnote{[Online] Available: \url{http://archive.ics.uci.edu/ml/machine-learning-databases/breast-cancer-wisconsin/}} 
and highlight the privacy-utility trade-off (PUT) in terms of diagnosis accuracy and loss function for different levels of privacy.  We also  compare our scheme with non-private  centralized setup, highlighting that by having 20 medical clients, diagnosis accuracy of 85\% can be achieved with privacy level of $\epsilon=20$, while the non-private  (and non-practical) centralized scheme achieves 95\% accuracy.}

In the following, we briefly review three state-of-the-art (SoTA) papers that are related to our work.    

\textit{\textbf{Related Works:}} 
In the context of  distributed learning protocols to privatize collaborative training, a related work \cite{Dopamine}  assumed a  system  in which  patients trust their local medical clinics, while clinics are assumed to non-malicious. 
%The parameter  server in their model is assumed as honest-but-curious.
This assumption does not necessarily hold in the realm of metaverse with a vast number  of AI-generated digital  healthcare entities,  without maintaining  any proper trust/authenticity guarantees \cite{medical_deepfake, synthetic}.    
%Finally, hospitals and the server do not trust any other third parties. 
We also note that the authors in \cite{Dopamine} only show  empirical results in terms of accuracy, but their paper lacks  any study on the PUT. 
%{Moreover, the work of  \cite{Dopamine} only showed empirical results by simulations, but lacked theoretical analysis on the FL system, such as tradeoff between privacy, convergence performance, and convergence rate.}  

An alternative approach for safeguarding the privacy of distributed learning algorithms is based on secure aggregation (SecAgg) methods with rigorous cryptographic guarantees \cite{SecAgg}. The system model assumes clients who can collude with each other to infer the local models of other users. 
%considering up to $T$ colluding entities who try to infer the local models of other users.
Formal privacy guarantees are provided from an information-theoretic perspective based on mutual information metric. However, SecAgg schemes cannot be considered as stand-alone solutions for privatizing  distributed learning algorithms due to the following facts: 
1) The main bottleneck of SecAgg is the client computation bottleneck. That is,   
%``client computation'', as 
the required  computation load on  each entity  scales linearly with the total number of participating nodes in the learning process.    
2) According to \cite{ai.googleblog}, while SecAgg tries to minimize data exposure, it cannot necessarily guarantee not to reveal  any information to malicious actors. Then DP  comes in as a mathematical framework with formal guarantees about privacy of 
%to  set  a limit on the  individual's influence on the outcome of a 
computations, such as AI/ML algorithms.  
Accordingly, we consider in this work the concept of DP as a widely-adopted technique in  data science and computer science literature. 
3) Besides, there might exist man-in-the-middle malicious parties who try to manipulate the public key exchange of SecAgg scheme \cite{WSKG-letter}.

We finally mention that the promising approach of employing ``{randomization mechanisms}'' can also bring about achieving \emph{formal guarantees}  on   DP   and convergence performance \cite{FL_Quek}. However, the  exact closed-form expressions for convergence bounds directly depend on the specific format of the chosen loss functions  and optimization algorithms.  
%Generally speaking,  DP-guaranteed distributed learning protocol  designers have to deal with the trade-off between privacy and utility of the employed  algorithms, which is known as PUT \cite{FL_Quek}.    

\subsection{Paper Organization and Notations} 
{The remainder of this paper is organized as follows.  Following the general format of papers in computer science literature, we first start by introducing  the definition of differential privacy and its mathematical framework in Section \ref{sec:Preliminaries} together with the corresponding interpretation in the metaverse.  We also provide a brief example to mathematically motivate the readers regarding the privacy risks of distributed learning algorithms.    In Section \ref{sec:FLDP}, our proposed global DP protocol for distributed learning is introduced with the mathematical models and algorithms. Then our scheme is evaluated in Section \ref{Sec:Exm_Res} using a real-world dataset. Finally, Section \ref{Sec:Concl} concludes our paper. 
}

\textit{Notations:} Vectors are represented by bold lower-case symbols, while sets are denoted by calligraphic symbols.  $\bm 0$ and $\bf I$ respectively show all-zero vector and identity matrix of the corresponding size. Moreover, $[N]$,  (with $N$ as integer) denotes the set of all integer values from $1$ to $N$. Operator $\mathds{1}_{x}$ outputs $1$ if $x>0$, and $0$ otherwise.

%==================================================================================================================================================
\section{Mathematical Background}\label{sec:Preliminaries}
In this section, we first  provide a brief example to mathematically motivate the readers regarding the privacy risks of distributed learning algorithms.    
Next, we provide the preliminaries and mathematical  background knowledge on the concept of client-level distributed DP, together with its  corresponding interpretations in the context of metaverse.   

Although distributed learning algorithms maintain locally-trained models,  sensitive information can still be inferred  if one analyzes the model parameters that are shared with  ML/AI entities \cite{model_inversion, backdoor}.  
%If the model updates   are inspected by bad actors, participant  users’ privacy  would be threatened. 
Generally speaking, FL algorithms can be considered as  ``privacy-aware'' methods for the metaverse-enabled distributed AI/ML healthcare algorithms. This is because the personal data protection is enhanced when  processed locally  by distributed metaverse systems. However, local model updates in distributed clinical  AI/ML algorithms carry extensive information about the  medical datasets owned by the e-health  agents.  Due to this fundamental fact,  malicious metaverse  participants can save the model parameters  and infer  private  attributes of patients' data. They might also be able to accurately recover the data  samples of users  \cite{model_inversion, backdoor}. 
		%Generally speaking, such attacks are typically known as inference attacks.  

\begin{figure}
		\centering
		\includegraphics
		[width=3.25in,height=1.8in,trim={0.0in 0.0in 0.0in  0.0in},clip]{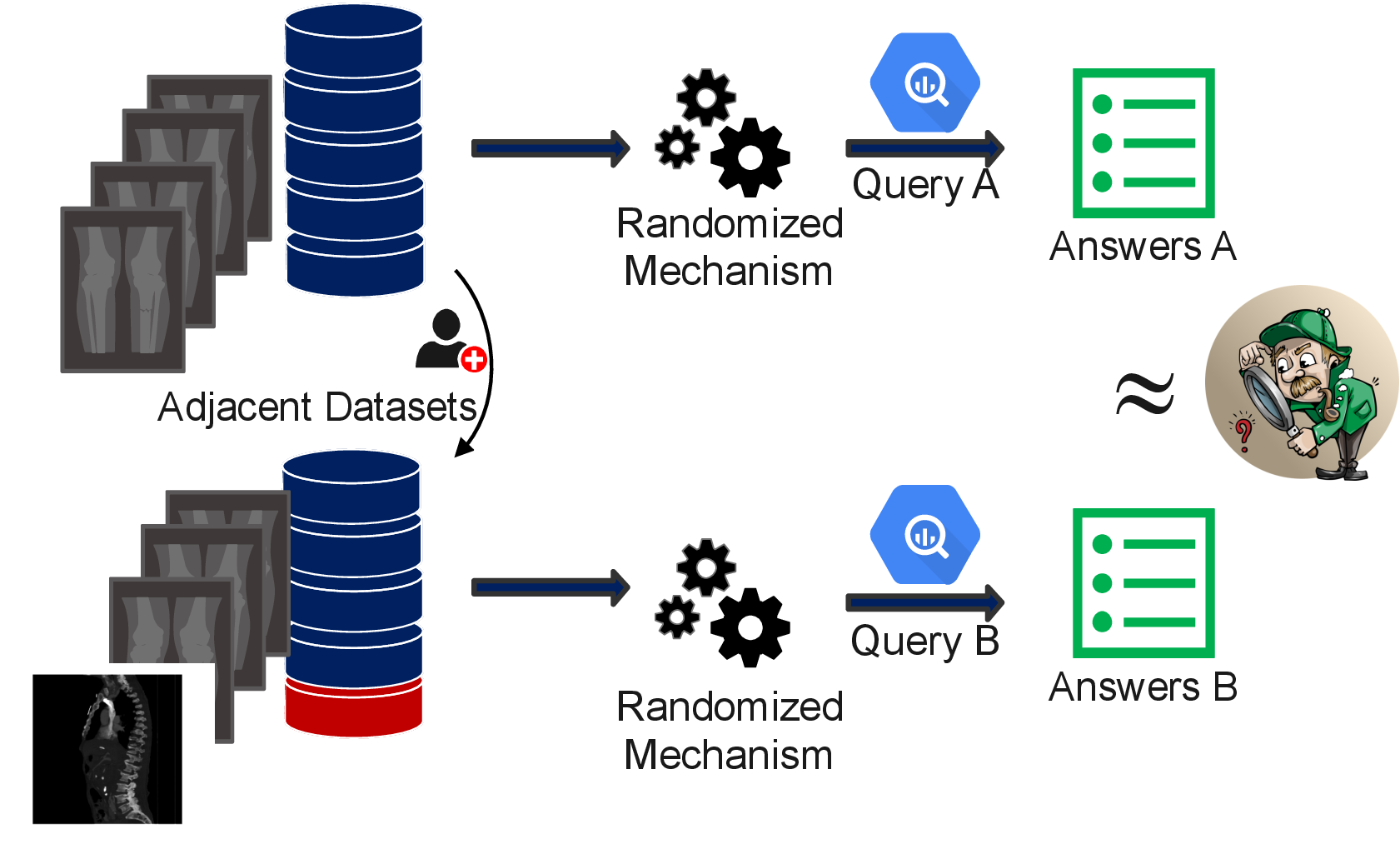}
		\vspace{0mm}\caption{The key idea of differential privacy.}
		\label{fig:DP}
		\vspace{-3mm}
	\end{figure}

\textit{A Mathematical Example:}    {To better understand the  privacy risk of distributed learning models, consider a simple ML  task with least-squares loss function  $\mathcal{L}_i(\boldsymbol{w}_i; \boldsymbol{x}_i,y_i) = | y_i - \boldsymbol{w}_i^{\sf T} \boldsymbol{x}_i |^2$ at the $i$-th learning entity within the FL framework, where input data, output label,  and model are denoted by  $\boldsymbol{x}_i$, $y_i$,  and  $\boldsymbol{w}_i$, respectively.  The stochastic gradient $\boldsymbol{g}_i$ of the corresponding loss function  would be
		\begin{align}\label{eq:example}
			\boldsymbol{g}_i  \overset{\Delta}{=}	\nabla \mathcal{L}_i(\boldsymbol{w}_i; \boldsymbol{x}_i,y_i) = \boldsymbol{x}_i (y_i - \boldsymbol{w}_i^{\sf T} \boldsymbol{x}_i). 
		\end{align}
  $\boldsymbol{g}_i$ is clearly proportional to the raw data $\boldsymbol{x}_i$. Hence, sharing the local  gradient updates allows the inference of users' datasets by malicious actors in the metaverse network.} 
	
	%As a countermeasure to privacy  leakage  and model inversions \cite{model_inversion},  a promising  approach is to employ 
	%%randomized mechanisms to fulfill privacy guarantees, where   a promising framework in this context is 
	%differential privacy (DP) \cite{FL_Quek}. 
	%Each learning entity  perturbs its model parameters,  and 
	%sends the randomized version to the aggregator node, to  enhance the privacy against both the malicious clients and curious servers. 
	%Notably,  there exist a trade-off  between the privacy protection level and the inference  performance, i.e., a better performance leads to a lower level of privacy, where a comprehensive trade-off analysis for differentially private FL framework can be found in \cite{FL_Quek}.

We now  introduce the definition of DP and its mathematical framework.  This is considered as a promising countermeasure to privacy  leakage  and model inversions \cite{AML_tomg, FL_Quek}.

\subsection{$(\epsilon, \delta)$-Differential Privacy}
Generally speaking,  $(\epsilon, \delta)$-DP provides a formal guarantee for preserving the privacy of distributed data processing algorithms.   An inherently  promising approach for preventing privacy leakage is to  employ perturbations via  ``{randomization mechanisms}'',  leading to achieving  \emph{formal guarantees} on DP.  
Mathematically speaking,  we have the following definition.  
\begin{definition}
\label{def:DP}$(\epsilon, \delta)$-DP:
A randomized mechanism  $\mathcal M: \mathcal{X}\rightarrow \mathcal{R}$ with domain $\mathcal{X}$ and range $\mathcal{R}$ guarantees  $(\epsilon, \delta)$-DP,
if for all measurable sets $\mathcal S\subseteq \mathcal{R}$ and for any two adjacent databases $\mathcal D_i, \mathcal D_i'\in \mathcal{X}$, the following condition holds 
\begin{equation}\label{equ:Differential privacy}
\mathbb{P}[\mathcal M(\mathcal D_i)\in \mathcal S]\leq \exp({\epsilon})\hspace{1mm}\mathbb{P}[\mathcal M(\mathcal D_i')\in \mathcal S]+\delta.
\end{equation}
\end{definition}
In this definition, $\epsilon > 0$ accounts  for  the  distinguishable bound on the set of  outputs corresponding to  neighboring datasets $\mathcal D_i$ and  $\mathcal D_i'$ in a database, and $\delta$ accounts for  the event that the ratio of the probabilities for two adjacent datasets $\mathcal D_i$ and $\mathcal D_i'$ cannot be bounded by $\exp({\epsilon})$ after applying a privacy preserving mechanism. With an arbitrarily given $\delta$, the larger $\epsilon$ is, the clearer the distinguishability of neighboring datasets would be,  increasing  the risk of privacy violation. 

Typically in practice, a randomized mechanism  $\mathcal{M}(\cdot)$  ensures differential privacy by adding calibrated ``mix-up''  noise to the output of a deterministic function, i.e.,  $\mathcal{M}(\bm x) = f(\bm x) +\mathsf{n}$.\footnote{For  the case of boolean attributes, the randomized mechanism can be formulated by $\mathcal{M}(\bm x) = \mathcal{R}\left(f(\bm x)\right)$, where $\mathcal{R}(\cdot)$ stands for a randomized response, e.g., in the form of a  biased coin flip.} 
		Lower $\epsilon$ values correspond to higher level of noise, making it more difficult  to distinguish outputs, thus, enforcing the privacy. 
		%In addition to $\epsilon$, 
		The required noise is also affected by the sensitivity $\Delta f$ of the real-valued function $f(\cdot)$, which quantifies the maximum difference between the function’ outputs corresponding to ${\cal D}_i$ and ${\cal D}_i^\prime$. The sensitivity is defined as follows. 
\begin{align}\label{eq:sensitivity} 
\Delta f =\max_{\mathcal D_i, \mathcal D_i'}{\Vert f(\mathcal D_i)-f(\mathcal D_i')\Vert}. 
\end{align}  
According to \cite{Gaussian}, in the scenario  of working with   numerical data, a Gaussian mechanism  can be applied  for guaranteeing  $(\epsilon, \delta)$-DP. Following the framework of \cite{FL_Quek}, to ensure that the mix-up noise sample $\sf n$  preserves $(\epsilon, \delta)$-DP,  we should have 
\begin{align}\label{eq:noise}
 \mathsf{n}\sim \mathcal{N}(0,\sigma^{2}), \quad \sigma\geq c\Delta f/\epsilon, 
\end{align}
where $\mathcal N$ represents the Gaussian distribution and $\sigma$ sets the mix-up scale. Moreover, the constant $c$ should satisfy  $c\ge\sqrt{2\ln(1.25/\delta)}$ \cite{Gaussian, FL_Quek}. 
%for $\epsilon \in (0,1)$.   

\subsection{Interpretation in The Metaverse Context}
%To deal with the unprecedented  data-centric privacy risks in the metaverse platforms, the idea of “incognito mode” has recently been developed for the metaverse as a novel and promising solution  for safeguarding VR users' privacy \cite{Vivek2}.    
%	To realize the incognito mode in the metaverse,  client-level  $\epsilon$-differential privacy framework is proposed by the researchers which has been  shown to be capable of protecting a wide variety  of sensitive data attributes in VR applications.  
In the context of metaverse, according to the incognito mode framework \cite{Vivek2},   \emph{the key idea of client-level DP for the metaverse would be to ensure that the observable attribute profile of a metaverse entity/user significantly overlaps with that of at least several other entities/users/avatars, making it intractable to accurately  determine personal identities.} 
%	Differentially-private algorithms enhance the privacy level  by adding randomized noise to an analysis, in a way that the output does not differ    with or without the presence of a specific data subject. 
 In the context of distributed learning, DP can be  considered as a mathematical framework, aiming to set  a limit on an individual's contribution to the output of a computation process, e.g.,  an  ML/AI algorithm. This is achieved  by bounding the contribution of any client and adding perturbation in the form of ``mix-up'' noise during  training
 %to produce a probability distribution over output models 
 \cite{ai.googleblog}. DP maintains  the  parameter ($\epsilon$) that helps  quantify  the amount of  distribution change when one adds or removes the training samples of any individual intelligent entity.    
 Intuitively and by invoking \eqref{equ:Differential privacy},   DP  provides a mathematical framework such that the information obtained by a ``hypothetical adversary'' about the output/response of a	given function/query is bounded. 
	The key idea of DP is also shown in Fig. \ref{fig:DP} for the sake of visual clarification.

%-----------------------------------------------------------------------------

%================================================================================================
\section{Proposed Model: Distributed Differential Privacy for Metaverse Healthcare}\label{sec:FLDP} 
%----------------------------------------------------------------------------------------------------------------------
	
	\begin{figure}
		\centering
		\includegraphics
		[width=3.4in,height=3.0in,trim={0.0in 0.0in 0.0in  0.0in},clip]{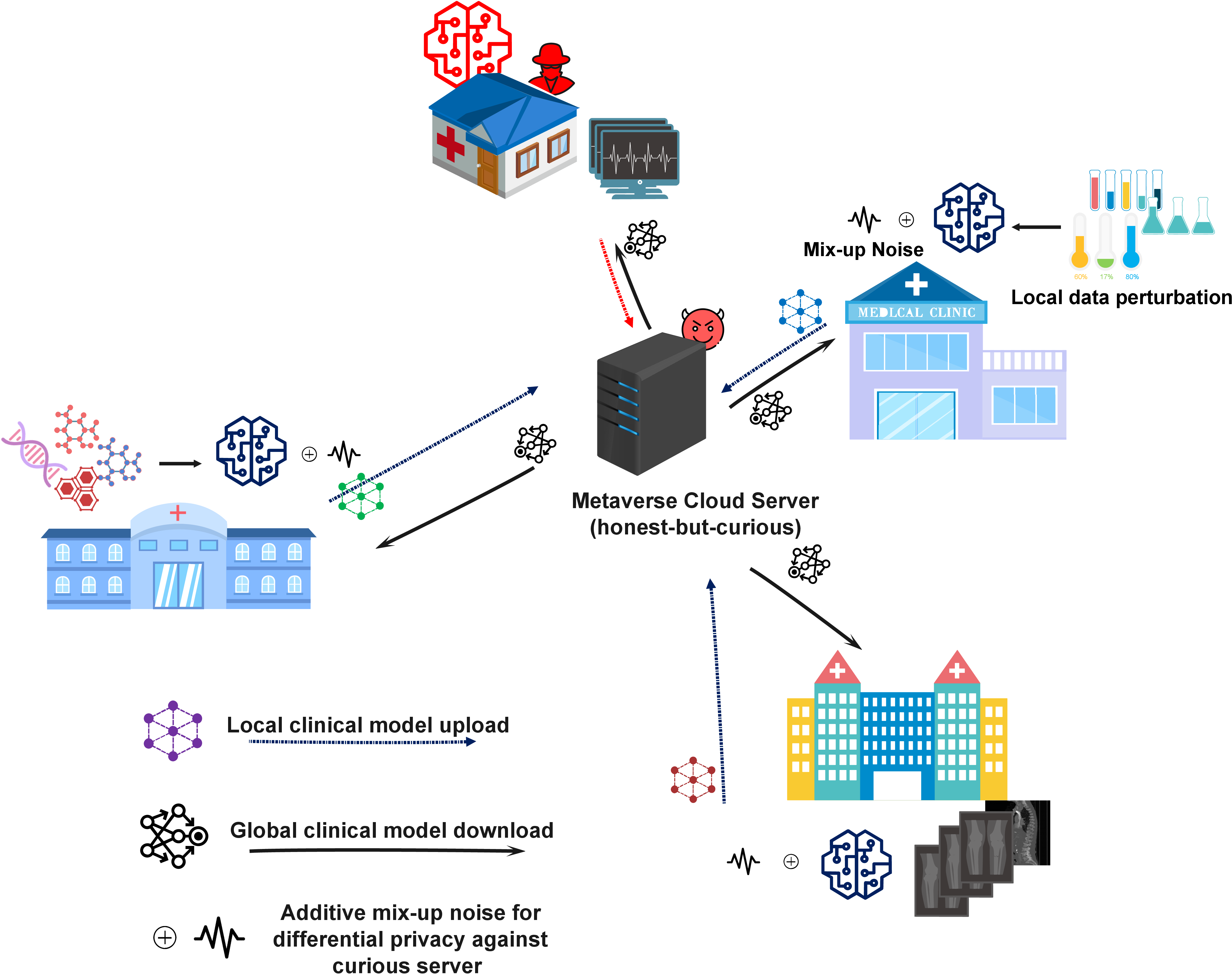}
		\vspace{2mm}\caption{General schematic of the studied system model in Metaverse-enabled distributed clinical AI: Mix-up noise is added to the local clinical models to adaptively provide distributed privacy for ML/AI-based  metaverse healthcare services.}
		\label{fig:backdoor}
	\end{figure} 

%\subsection{System Overview}\label{sec:Federated Learning}
Consider a  distributed metaverse-enabled digital healthcare platform, in which every virtual clinic or medical center in the metaverse, denoted by $\mathcal{V}_i$,  $i\in \{1, 2,\cdots, V\}$, aims to carry out a medical diagnosis task with the help of ML/AI algorithms, without disclosing the medical data of its metaverse patients to other virtual clinics, avatars, and servers. A general description  of the platform is depicted in Fig. \ref{fig:backdoor}.
Each virtual hospital or medical centre  maintains its own medical  database $\mathcal D_i$ for $i\in \{1, 2,\cdots, V\}$. With the aid of a facilitator entity, such as a metaverse server in the backbone of the metaverse network, 
each ML/AI-based virtual clinic tries to find a vector $\bm {w}_i$ corresponding to its medical AI diagnostic model.   Mathematically speaking, we have 
\begin{align}\label{localtrain}
\bm{w}_{i}&=\arg\min\limits_{\bm{w}} \hspace{1mm}{\mathcal{L}_{i}(\bm{w}, \mathcal D_i)}\nonumber \\
&=\frac{1}{\vert \mathcal D_i \vert}\sum_{j = 1}^{\vert \mathcal D_i \vert}\arg\min\limits_{\bm{w}}\hspace{1mm}{\mathcal{L}_{i}(\bm{w}, \mathcal D_{i,j})}, 
%\quad i\in \{1, 2,\cdots, V\}, \quad j\in\{1,\cdots,|\mathcal{D}_i|\} 
\end{align}
where $\mathcal{L}_{i}(\cdot, \cdot)$ stands for the medical  loss function of the $i$-th virtual clinic, $i\in [V]$. $\mathcal{L}_{i}(\cdot, \cdot)$  is calculated  based on  local empirical risks. Moreover, $\mathcal D_{i,j}$ is the $j$-th sample in  medical database $\mathcal D_i$, $j \in  \big[|\mathcal{D}_i|\big]$. 
To facilitate the distributed learning process, a metaverse  server takes  care of aggregating all  the medical models. 
%sent from the $V$ virtual clinics. 
This model sharing can  be  realized by utilizing  the concept of decentralized networking in the metaverse
%in the backbone of the metaverse systems
based on the so-called Web 3.0 protocol \cite{accenture}. 
The server is considered to be honest-but-curious, i.e.,  it follows the learning protocols while it can also try to obtain  information  about users from data it receives.   
Mathematically speaking, we have the following general expression for the aggregated medical model   
\begin{equation}\label{equ:Aggregating}
\bm{w} =\sum_{i=1}^{V}{\pi_{i}\bm{w}_{i}},
\end{equation}
where $\pi_{i} = {\vert \mathcal D_i\vert}/{\vert \mathcal D\vert}$ stands for  the contribution ratio of each virtual clinic in the metaverse,   
%with $\sum_{i=1}^{N}{\pi_{i}}=1$, 
and $\vert \mathcal D\vert = \sum_{i=1}^{V}{\vert \mathcal D_{i}\vert}$ shows the size of the entire  medical data samples. Accordingly,  the generic framework for  obtaining a medical ML/AI model via collaborative learning is to solve the following  problem. 
\begin{equation}\label{equ:Global objective function}
\bm{w}^{*}=\arg\min_{\bm{w}}{\sum_{i=1}^{V}{\pi_{i}\hspace{1mm} \mathcal{L}_{i}(\bm{w})}}. 
\end{equation}
According to the FL literature,  the $V$ virtual clinics having similar data structure  are able to collaboratively learn the ML medical model with the help of a medical cloud server (a.k.a parameter server).  This is achievable after running the training algorithm  for a sufficient number of rounds for local training and model exchange. Then as proved in \cite{FL_Quek}, the solution to the optimization problem in  \eqref{equ:Global objective function} can potentially converge to that of the global optimal learning model. 

In the context of  distributed learning protocols, metaverse patients and medical clinics  cannot necessarily trust other virtual  clinics. This is due to the large number  of AI-generated digital  healthcare entities,  without maintaining  any proper trust/authenticity guarantees \cite{medical_deepfake, synthetic}. Besides, the distributed  cloud servers might be honest-but-curious with respect to  their communication and computation protocols.   
As already shown mathematically in \eqref{eq:example}, although distributed learning algorithms render local training,  sensitive information can still be obtained   if one  analyzes  the model parameters that are collaboratively shared with the server \cite{model_inversion, backdoor}.  
Therefore,  inspired by the idea of ``incognito mode'' metaverse \cite{Vivek2}, 
%based on client-level  $\epsilon$-differential privacy, in this paper  
we propose  \emph{distributed differential privacy} for intelligent metaverse healthcare systems.   
%Each intelligent healthcare  party perturbs its model parameters,  and  sends the randomized version to the aggregator node, to  enhance the privacy against both the malicious actors  and curious servers.
While in \cite{Vivek2} the mix-up noise is directly applied to metaverse users' data,  in our scheme, we  generalize the concept of  ``client-level'' DP  proposed in  \cite{Vivek2}, and realize  \emph{global differential privacy} for distributed learning-based  metaverse healthcare systems.  
More specifically, in our scheme,  randomized mechanism introduced in  \eqref{equ:Differential privacy} is realized in  the form of ``mix-up'' noise which is applied to the federated clinical ML/AI models---each of the virtual clinics  perturbs its clinical model  parameters by adding mix-up noises before sharing them with the parameter server for aggregation. Then in the aggregation phase, this perturbation is adaptively fine-tuned to meet the privacy requirements  during global model update. Our global  $(\epsilon, \delta)$-DP framework can provide an adjustable level of  privacy against both the malicious actors and honest-but-curious metaverse servers.

\subsection*{Protocol for Global Differential Privacy} 
%Here, we define a global $(\epsilon, \delta)$-DP requirement for the collaborative medical learning in the metaverse.  
In this subsection, we describe in details the protocol that should be run in  a distributed metaverse  healthcare system  to realize global distributed DP. 

At the beginning of the differentially-private distributed learning protocol, the cloud server broadcasts the configurations regarding the required level of privacy $(\epsilon, \delta)$ together with an  initial value for the  global parameter $\bm{w}^{(0)}$.  During the  $t$-th round of collaborative training ($t>0$),  each virtual clinic   trains its medical model based on \eqref{localtrain}, using its local medical database $\mathcal{D}_i$, $i\in [V]$. 
%with preset termination conditions. 
Once  the local training is completed,  the $i$-th clinic  perturbs its model by first clipping the corresponding vector and then  adding mix-up noises to the locally-trained model parameter $\bm{w}^{(t)}_i$. 
To elaborate,   the clipping gradient technique is applied to the local models such  that $\Vert\bm{w}^t_{i}\Vert \leq B$, where 
%$\bm{w}_{i}$ denotes training parameters from the $i$-th client without perturbation and
$B$ denotes the  clipping threshold for  bounding  the model parameters.  
Accordingly, the bounded  version of the medical model vector at the $i$-th clinic can be expressed  as  
\begin{align}\label{eq:clipping}
\bm{w}^{(t)}_{i, \hspace{1mm}\mathsf{clipped}} = \bm{w}^{(t)}_{i}/\max\left(1,\frac{\Vert\bm{w}^{(t)}_{i}\Vert}{B}\right). 
\end{align}
Invoking \eqref{eq:sensitivity} and applying it to  \eqref{localtrain} and \eqref{eq:clipping}, the corresponding model sensitivity $\bm{w}_{i}^{(t)}$ of the $i$-th client  can be formulated as 
\begin{multline}\label{SensitivityforUP}
S_{i}^{(t)} =
%\max_{\mathcal D_i, \mathcal D_i'}{\Vert \bm{w}_{i, \mathcal{D}_i}^{(t)} - \bm{w}_{i, \mathcal{D'}_i}^{(t)}\Vert}\\
\max_{\mathcal D_i, \mathcal D_i'}\left\Vert \frac{1}{\vert \mathcal D_i \vert}\sum_{j = 1}^{\vert \mathcal D_i \vert}\arg\min\limits_{\bm{w}}{\mathcal{L}_{i}(\bm{w}, \mathcal D_{i,j})}\right.\\
\left.-\frac{1}{\vert \mathcal D_i' \vert}\sum_{j = 1}^{\vert \mathcal D_i' \vert}\arg\min\limits_{\bm{w}}{\mathcal{L}_{i}(\bm{w}, \mathcal D_{i,j}')}\right\Vert= \frac{2B}{\vert \mathcal D_i \vert},
\end{multline}
where $\mathcal D_i'$ is an adjacent dataset to $\mathcal D_i$ that  maintains  the same size but differs by one sample  (according to the definition  of DP  framework proposed in Definition \ref{def:DP}). Moreover,  $\mathcal D_{i,j}'$ is the $j$-th sample in $\mathcal D_i'$.
Accordingly, a  global sensitivity for the model sharing phase  can be expressed as 
$S_{\sf sharing} = \max\limits_{i\in [V]}\left\{ {S}_{i}^{(t)}\right\}$. 
%\begin{equation}
%{\bm s}^{(t)}_{\sf{sharing}}\triangleq \max\left\{\Delta \bm{w}_{i}^{(t)}\right\},\quad\forall i \in \{1,\cdots, V\}.
%\end{equation}
Invoking \ref{SensitivityforUP} and assuming  that the  minimum size of the local medical  datasets owned by  virtual clients is $m$, one should satisfy $S_{\sf{sharing}} = \frac{2B}{m}$. 

With the aim of guaranteeing  global  $(\epsilon, \delta)$-DP for all of the distributed virtual clients during the $t$-th round of medical model sharing,  the mix-up noise scale  must be  chosen as $\sigma_{1}= c S_{\sf{sharing}}/\epsilon$, as introduced in \eqref{eq:noise}.
Hence, having  $T$ rounds of collaborative model sharing and aggregation, we consider  that we have at most $E$ ($E\leq T$)  exposures of  model to malicious entities during local model upload \cite{FL_Quek}.  
%and T exposures of aggregated parameters
%in the downlink, where T is the number of aggregation times.
Thus, one should choose $\sigma_{1}= cE S_{\sf{sharing}}/\epsilon$. The multiplication by $E$ is because in the Gaussian randomized mechanism, the mix-up noises follow Gaussian distribution, and  are uncorrelated with respect to different rounds of applying the Gaussian mechanism.  
%This also shows the the standard deviation of the additive Gaussian noise. 
Mathematically speaking,  the process of medical model mix-up can be formulated as follows.  
\begin{align}\label{eq:model_perturb} 
\widetilde{\bm{w}}^{(t)}_i  = \bm{w}^{(t)}_{i, \hspace{1mm}\mathsf{clipped}}  + \bm{n}_i^{(t)}, \quad  \bm{n}_i^{(t)} \sim \mathcal{N}(\bm{0}, \frac{cT S_{\sf{sharing}}}{\epsilon}\bf I).  
\end{align}

%\textcolor{red}{The next step is to apply a randomized mechanism, i.e.,
%adding mix-up noise at the client side first and then decide
%whether or not to add noises at server to satisfy the $(\epsilon, \delta)$-DP
%criterion in the downlink channel. before uploading  the intentionally  mixed-up model vector  $\widetilde{\bm{w}}^{(t)}_i$ to the metaverse server.}
Medical clients then share their models with the metaverse cloud server, where  all of the local model vectors  
%denoted by $\widetilde{\mathbf{W}} = \{\widetilde{\mathbf{w}}_{1}, \ldots,\widetilde{\mathbf{w}}_{N}\}$, 
are received by the server. 
\emph{Notably, since we employed model perturbations in \ref{eq:clipping}--\eqref{eq:model_perturb} by exactly following the steps of DP framework,  according to \eqref{equ:Differential privacy}, any malicious actor in the metaverse (including an honest-but-curious server) cannot infer private information  of local clients from  the medical  models uploaded to the server.}  

%The metaverse cloud server then aggregates the  intentionally  mixed-up model vector of all medical clients $\widetilde{\bm{w}}^{(t)}_i$, $i \in \{1, \cdots, V\}$ to  find a global model update ${\bm{w}}^{(t)}$.
At the metaverse server,  an  aggregation process can be generally formulated by 
%for $\mathcal D_i$ can be expressed as
\begin{align}\label{eq:Agg}
%s_{\text{D}}^{\mathcal D_i}\triangleq 
\bm{w}^{(t)}&=\sum_{i \in V}\pi_i \widetilde{\bm{w}}^{(t)}_i 
\nonumber \\
&= \sum_{i \in V}\pi_i {\bm{w}}^{(t)}_{i, \hspace{1mm} \sf clipped} + \sum_{i \in V}\pi_i {\bm{n}}^{(t)}_i,  
\end{align}
where $\bm{w}^{(t)}$ is the aggregated medical model in the $t$-th round of distributed learning.  
%supposed to be sent back to the clients for model update.  

After the model aggregation at the metaverse cloud server, $\bm{w}^{(t)}$ is supposed to be broadcast  to the medical clients for model update.  
This again incurs privacy risks to  the network, and sensitive information about individual health entities  can be revealed from $\bm{w}^{(t)}$,  if overheard by potential totally-passive malicious actors in the metaverse network.  Therefore, we  fine-tune our distributed learning model in terms of privacy preservation. To elaborate, privacy guarantee with respect to the global model sharing  is ensured by applying the randomized Gaussian mechanism on the global model $\bm{w}^{(t)}$  before sending the models back to the virtual medical clinics.  
Following the generic framework of DP in Definition \ref{def:DP}, one first needs to obtain  the sensitivity of $\bm{w}^{(t)}$ with respect to each of the datasets $\mathcal{D}_i$ for $i \in [V]$. We denote  the corresponding sensitivity by $S_{\bm w, \mathcal{D}_i}^{(t)}$. Accordingly,  by  calculating  $S_{\bm w, \mathcal{D}_i}^{(t)} \triangleq \max_{\mathcal D_i, \mathcal D_i'}\left\Vert {\bm{w}}^{(t)}(\mathcal{D}_i) - {\bm{w}}^{(t)}(\mathcal{D'}_i) \right \Vert$, one can obtain the sensitivity of aggregation phase \eqref{eq:Agg} for every dataset $\mathcal D_i$, $i \in [V]$, with a similar approach to \eqref{SensitivityforUP},  
%after the aggregation operation $s_{\emph{D}}^{\mathcal D_i}$ 
as given below 
\begin{equation}\label{equ:SensitivityforDL}
\begin{aligned}
S_{\bm w, \mathcal{D}_i}^{(t)} = \frac{2B\pi_{i}}{m}.
\end{aligned}
\end{equation}
This can be obtained in a straightforward manner, where we omit repeating the details.  We simply hint that the notations $\bm{w}^{(t)}(\mathcal{D}_i)$ and $\bm{w}^{(t)}(\mathcal{D'}_i)$ indicate that the $i$-th model is trained with dataset $\mathcal{D}_i$ and $\mathcal{D'}_i$, respectively.  
Then the global sensitivity for model broadcast can be defined as $S_{\sf broadcast} = \max\limits_{i\in [V]}\left\{ S_{\bm w, \mathcal{D}_i}^{(t)}\right\}=
\max\limits_{i\in [V]}\left\{{2B\pi_i}/{m} \right\}$. The minimum sensitivity is achieved when we have  $\pi_{i} = 1/V, \forall i \in [V]$.   

%To achieve a small global sensitivity in the downlink channel which is defined by
%\begin{equation}
%\Delta s_{\emph{D}}\triangleq \max\left\{ \Delta s_{\emph{D}}^{\mathcal D_i}\right\}=\max\left\{ \frac{2C\pi_{i}}{m}\right\},\quad \forall i,
%\end{equation}
%the ideal condition is that all the clients should use the same size of local datasets for training, i.e., $\pi_{i}=1/N$. to  obtain the optimal value of the sensitivity $\Delta s_{\text{D}}$.

 As already mentioned,  our distributed DP protocol is based on applying  a randomized mechanism at the local models,  and then fine-tuning that mechanism at the server side, e.g., by adding additional mix-up noise,  aiming  to guarantee the global $(\epsilon, \delta)$-DP among any malicious or curious actor in the network.  In what follows,  we continue by providing  the design guidelines for ensuring the global distributed DP.

%To ensure a global $(\epsilon, \delta)$-DP in the uplink channels, the standard deviation of additive noises in client sides can be set to $\sigma_{\text{U}} = cL\Delta s_{\text{U}}/\epsilon$ due to the linear relation between $\epsilon$ and $\sigma_{\text{U}}$ with Gaussian mechanism, where $\Delta s_{\text{U}} = \frac{2C}{m}$ is the sensitivity for the aggregation operation and $m$ is the data size of each client.
%We then set the sample in the $i$-th local noise vector to a same distribution $n_{i}\sim\varphi(n)$ (i.i.d for all $i$) because each client is coincident with the same global $(\epsilon, \delta)$-DP.
Invoking the aggregation process in \eqref{eq:Agg},  
thanks to the characteristics of normal distributions, $\sum_{i \in V}\pi_i {\bm{n}}^{(t)}_i$ also follows the jointly normal  distribution.  This is because in our Gaussian randomization mechanism, we apply independent-and-identically-distributed (i.i.d.) mix-up noises to every local client  with  the  distribution given in \eqref{eq:model_perturb}. 
%Gaussian mechanism for $n_{i}$ with noise scale $\sigma_{\text{U}}$, the distribution of $\pi_{i}n_{i}$ is also Gaussian distribution. 
%To obtain a small sensitivity $\Delta s_{\text{D}}$, we set $\pi_{i} = 1/N$.
%Furthermore, the noise scale $\sigma_{\text{U}}/\sqrt{N}$ of the Gaussian distribution $\phi_{N}(n)$ can be calculated.
Following the DP framework based on equations \eqref{equ:Differential privacy}--\eqref{eq:noise} and applying it to \eqref{eq:Agg} and \eqref{equ:SensitivityforDL}, we conclude that  in order to ensure $(\epsilon, \delta)$-DP, the aggregated  mix-up scale level should satisfy 
  $\sigma = cT S_{\sf broadcast} /\epsilon$, where $S_{\sf broadcast} = 2B/(Vm)$.
Accordingly,  the mix-up level that might be required to be applied by  the cloud  server is calculated   as $\sigma_{2}=\sqrt{\Big(\sigma^{2}-{\sigma_{1}^{2}}/{V}\Big)}$, in which we have $\sigma_1 = 2BEc/(m\epsilon)$.  Hence,  the fine-tuning mix-up level can be rewritten  as   
\begin{algorithm}[t]
\caption{Global DP for Distributed  MetaHealth}
\label{alg:NbAFL}
\LinesNumbered
\texttt{Config:\hspace{2mm}}{($\epsilon$, $\delta$}), $\bm{w}^{(0)}$, and $T$ \\
\texttt{{Initialization:\hspace{0mm}}  $\bm{w}^{(0)}_{i} = \bm{w}^{(0)}$}, $\forall i \in [V]$, and $t = 1$\\
\While {$t \le T$}
{
\texttt{Local model training:}\\
\For {$i \in [V]$}
{
Local training based on \eqref{localtrain} %$\mathbf{w}^{(t)}_{i}$ as\\
%\quad\quad $\mathbf{w}^{(t)}_{i}=\arg\min\limits_{\mathbf{w}_{i}}{\left(F_{i}(\mathbf{w}_{i})+\frac{\mu}{2}\Vert \mathbf{w}_{i}- \mathbf{w}^{(t-1)}\Vert^{2}\right)}$
\\
Model perturbation based on \eqref{eq:model_perturb}
%Clip the local parameters $\mathbf{w}^{(t)}_{i} = \mathbf{w}^{(t)}_{i}/\max\left(1,\frac{\Vert\mathbf{w}^{(t)}_{i}\Vert}{C}\right)$\\
%Add noise and upload parameters $\widetilde{\mathbf{w}}^{(t)}_{i}=\mathbf{w}^{(t)}_{i}+\mathbf{n}^{(t)}_{i}$
\\
}
\texttt{Model sharing and aggregation: 
%(cloud server):
}
\\
Local models upload \& aggregation based on \eqref{eq:Agg} %$\mathbf{w}^{(t)}$ as\\
%\quad\quad $\mathbf{w}^{(t)} = \sum\limits_{i=1}^{N}{\pi_{i}\widetilde{\mathbf{w}}^{(t)}_{i}}$
\\
Global model perturbation based on \eqref{eq:server_noise} 
\\
Global model broadcast
%\quad\quad$\widetilde{\mathbf{w}}^{(t)}=\mathbf{w}^{(t)}+\mathbf{n}_{\text D}^{(t)}$
\\
\texttt{Local evaluation:}\\
\For {$i \in [V]$}
{
Locally Evaluate the models  using  $\widetilde{\bm{w}}^{(t)}$\\
}
$t \leftarrow t + 1$
}
\texttt{Output:\hspace{2mm}}{$\widetilde{\bm{w}}^{(T)}$}
\end{algorithm}

\begin{align}
\sigma_{2}= \frac{2Bc\sqrt{T^{2}-E^{2}V}}{Vm\epsilon} \mathds{1}_{(T-E\sqrt{V})}.
%\begin{cases}
%\frac{2Bc\sqrt{T^{2}-E^{2}V}}{Vm\epsilon} & T>E\sqrt{V},\\
%0&\text{Otherwise}.
%\end{cases}
\end{align}
Thus, the fine-tuning process at the cloud server can be formulated as 
\begin{align}\label{eq:server_noise}
\widetilde{\bm{w}}^{(t)} =  \begin{cases}
 \bm{w}^{(t)} + \bm{n}^{(t)}, \quad \bm{n}^{(t)} \sim \mathcal{N}(\bm{0}, \sigma_2\bf I) & T>E\sqrt{V},\\
{\bm{w}}^{(t)}&\text{Otherwise}.
\end{cases}
\end{align}

The overall  algorithm is summarized in \textbf{Algorithm~\ref{alg:NbAFL}}. 
The  algorithm stops after reaching a pre-configured  number $T$ of model update,  returning  $\widetilde{\bm{w}}^{(T)}$ as output.

\textit{\textbf{Remark}: }{\textit{Before providing the experimental results, we notice that applying the proposed distributed DP framework can also bring about resilience against the so-called backdoor attacks as well \cite{AML_tomg, backdoor}. In this context,  malicious actors may inject  mislabeled or manipulated data to mislead the global model,  which is known  as data poisoning attacks \cite{AML_tomg}. For example, when a patient is creating his/her avatar, a malicious entity  in the metaverse can modify its inferred  data, which can result in creating a different  avatar with different health-related attributes  than that of the actual patient. In our future work, we will apply our proposed framework to such cases.}}

\section{Evaluations}\label{Sec:Exm_Res}
In the evaluation section, we conduct experiment to numerically study the performance of the proposed globally-private protocol. Specifically,   we evaluate our scheme on a real-world dataset, i.e.,  breast cancer Wisconsin dataset (BCWD) \cite{Centralized_SVM_Wisconsin}.\footnote{[Online] Available: \url{http://archive.ics.uci.edu/ml/machine-learning-databases/breast-cancer-wisconsin/}}   
BCWD contains  683 non-missing data records, where  for each record, there are 9 discrete attributes with  values in the interval $[1,10]$. In this dataset,
458 records’ (65.5\%) predictions are benign and 241
records’ (34.5\%) predictions are malignant. 
The dataset is utilized to train a federated clinical model for  classification task, where we employ our learning model based on support vector machines (SVMs).  The experiments are carried out using Python 3.\footnote{\url{https://www.python.org/downloads/release/python-3913/}} 
%and real-world federated datasets. 
{We further note that our proposed scheme is not limited to any specific type of healthcare dataset, and can be easily applied to any other metaverse-enabled healthcare datasets \cite{dataset}, which will be studied in our future works.} 

\begin{figure}
	\vspace{0mm}
	\centering
	\includegraphics
 [width=3.0in,height=2.2in,trim={0.0in 0.1in 0.0in  0.0in},clip]{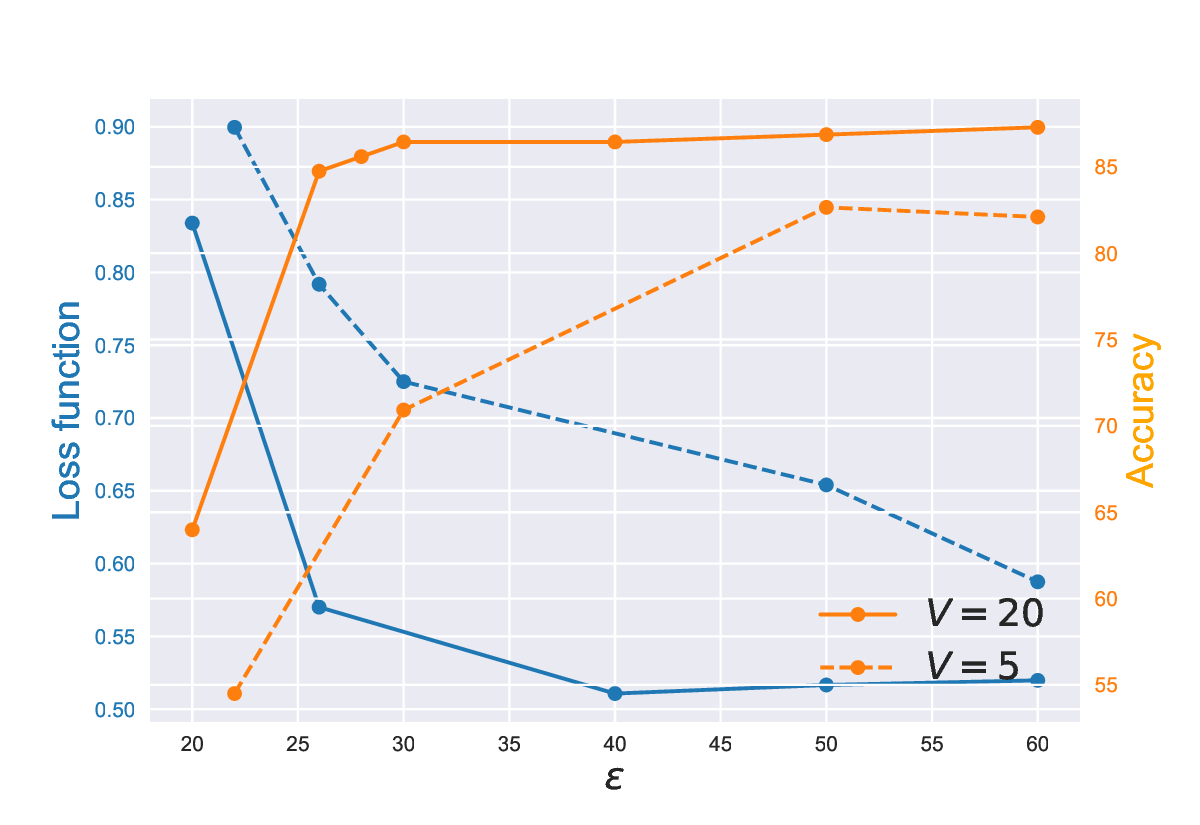}
\vspace{0mm}
\caption{PUT: Diagnosis accuracy and loss function vs. adjustable privacy level.}
	\label{fig:trade-off}
 \vspace{-3mm}
\end{figure} 

We conduct our experiments by investigating the privacy and accuracy of the proposed system over different values of privacy protection level $\epsilon$.  We also  study the effect of the number of participating  medical clients on the performance of system. 
%distributed learning protocol.
Furthermore, we compare our scheme with baselines of non-private setup and centralized settings.

 Fig. \ref{fig:trade-off}.   demonstrates  the   diagnosis accuracy and the corresponding  loss function values during learning process, considering  different levels of distributed privacy, denoted by $\epsilon$. As a reminder,  lower $\epsilon$ values indicate more stringent privacy requirements based on \eqref{equ:Differential privacy}, {where this is interpreted as the adjustable level of privacy in DP literature \cite{Vivek1, Vivek2, Vivek3, ai.googleblog, Dopamine}.}         
According to thew figure,  a  trade-off exists  between the inference  performance and the privacy protection level, which is generally known as privacy-utility trade-off (PUT) \cite{AE_paper}. The figure shows that by relaxing the privacy level requirement,  better diagnosis performance could be achieved. The good point here is that the PUT curves start to saturate. Hence, one can for example, choose the privacy protection level  $\epsilon=30$ (with $V = 20$ participating clinics) to achieve almost the same accuracy as $\epsilon=50$, while achieving better privacy protection.     
The figure further highlights that  increasing the number, $V$,  of medical clients participating in the distributed learning setup,  i.e., virtual medical clinics, higher diagnosis accuracy can be achieved (along with lower loss function values).  This happens due to the coordination and collaboration among more distributed  medical learning parties, sharing their ``learning insights'' with each other over time.

\begin{figure}
	\vspace{0mm}
	\centering
	\includegraphics
 [width=3.0in,height=2.2in,trim={0.0in 0.1in 0.0in  0.0in},clip]{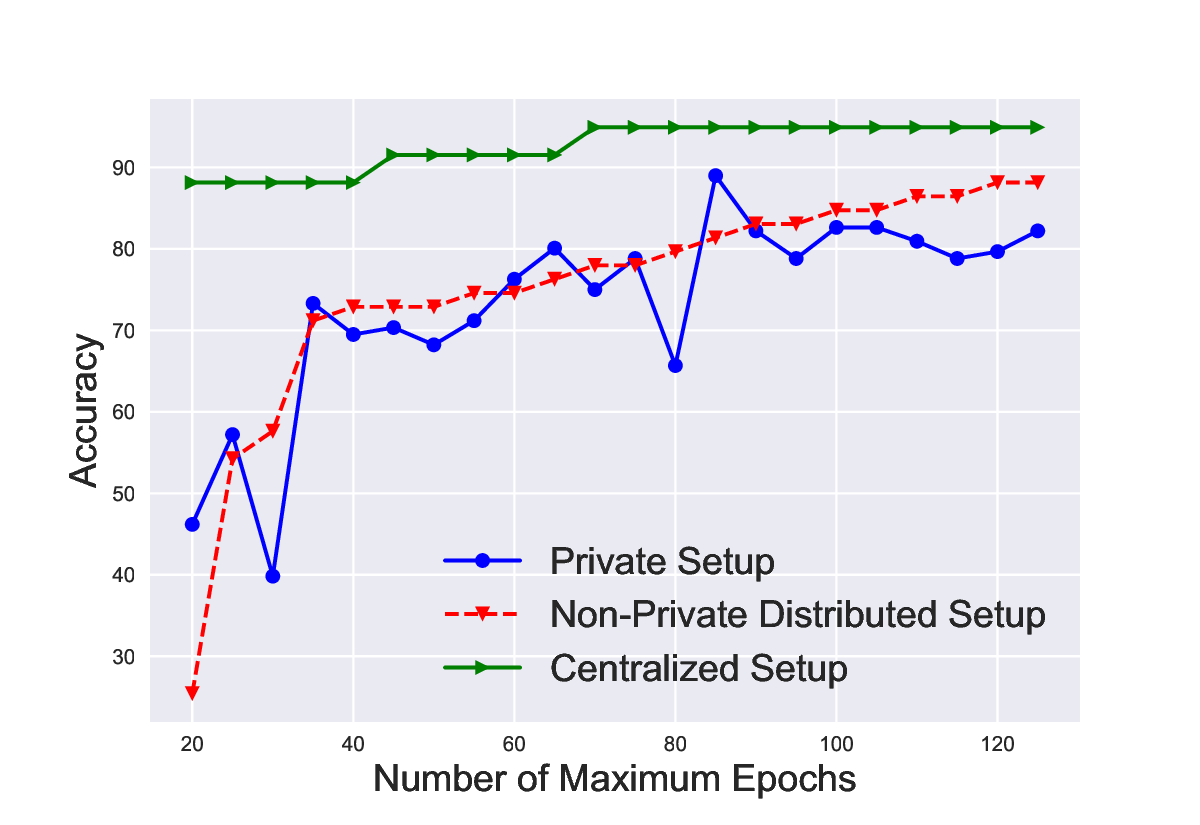}
\vspace{0mm}
\caption{Accuracy vs. the maximum number of training epochs for $\epsilon = 25$ and $V=20$:  Comparison with non-private and centralized settings.}
	\label{fig:baseline}
 \vspace{-3mm}
\end{figure} 

Fig. \ref{fig:baseline} illustrates the diagnosis accuracy versus the number of maximum training epochs.  The figure shows that by increasing the number of maximum training epochs, higher diagnosis accuracy can be achieved thanks to the more opportunity we give to the medical clients to gradually train their models.   
We also compare our  results with the corresponding  baselines of  having no privacy constraint, and also a centralized learning scheme.  Comparing our scheme with the non-private distributed setup, one can infer  that we can guarantee the global distributed privacy, while the accuracy does not fall that much. As we can see, the achievable accuracy for the private setup is only $5\%$ less than the non-private setup, while we already know that  the non-private setup is vulnerable to different privacy and security threats as already mentioned in our paper. 
The fluctuations seen in the accuracy values of the private setup (the blue curve) can be considered to be the result of having random mix-up noises injected into the local learning models of medical clients, making it more challenging for the local clinics to always maintain a progressive performance during all of the realizations of the employed learning protocol.   
Finally, comparing our scheme with the baseline of centralized setup, in which a central entity has access to all the medical datasets of different individuals and clinics (without any privacy mechanism), shows that if we apply our distributed DP mechanism,  we might loos less than  $10\%$ of accuracy, while we achieve  ``global DP'' in return.              

{Before concluding the paper, we would like to highlight that one of the main challenges in implementing our proposed approach  in a real-world practical setup would be to maintain a proper  balance between accuracy and privacy.  This might also vary between different tasks and datasets, and requires  fine-tuning the protocol in the deployment phase.} 

%=====================================================================================
\section{Conclusions}\label{Sec:Concl}
In this paper, we  proposed  global distributed DP for  intelligent  metaverse healthcare systems.    
%Each intelligent healthcare  party perturbs its model parameters,  and  sends the randomized version to the aggregator node, to  enhance the privacy against both the malicious actors  and curious servers.
Randomized mechanism in the form of ``mix-up'' noise was  applied to the federated clinical ML/AI models. Every virtual clinic perturbs its clinical model  vector  by adding mix-up noises before sharing them with the cloud server. Then in the aggregation phase, this perturbation is adaptively fine-tuned to meet the privacy requirements  during global model update. We provided global  $(\epsilon, \delta)$-DP framework against any malicious actors and honest-but-curious metaverse servers.   
We evaluated our scheme on BCWD dataset. We addressed  PUT in terms of diagnosis accuracy and loss function for different levels of privacy.  We also  compared our scheme with non-private  centralized setup.   
%=====================================================================================

\section{ACKNOWLEDGMENTS}
This research was supported by the College of Technological Innovation, Zayed University (ZU) under grant number RIF-23020.


\begin{thebibliography}{10}	
		%\bibitem{ericsson}
		%Y. Nezami, M. Dohler,  M. Shirazipour, and E.  Blomquist, ``\emph{What is the metaverse and why does it need 5G to succeed? The metaverse 5G relationship explained}.'' 
 
\bibitem{Metav1}
M. Letafati and S. Otoum, “On the privacy and security for e-health
services in the metaverse: An overview,” \emph{Ad Hoc Networks,} Aug. 2023,
https://doi.org/10.1016/j.adhoc.2023.103262.
%V. Ahsani, A. Rahimi, M. Letafati, and B. H. Khalaj, ``Unlocking Metaverse-as-a-Service The three pillars to watch: Privacy and security, edge Computing, and blockchain,''  \emph{arXiv:2301.01221v2}, Jan. 2023.   

\bibitem{Metav2}
M. Letafati and S. Otoum, ``Digital healthcare in the metaverse: Insights into privacy and security,'' 
\emph{arXiv:2308.04438v2}, Aug. 2023. [Online]. Available: https://arxiv.org/abs/2308.04438v2.


\bibitem{Moayad_BC}
M. Aloqaily, O. Bouachir, F. Karray, I. A. Ridhawi and A. E. Saddik, ``Integrating digital twin and advanced intelligent technologies to realize the metaverse,''  \emph{IEEE Consumer Electronics Magazine}, Oct. 2022. 
  
		\bibitem{Moayad_DT}
		M. Aloqaily, O. Bouachir, and F. Karraya, ``Digital twin for healthcare immersive services:
		Fundamentals, architectures, and open issues,'' in \emph{Digital Twin for
		Healthcare: Permissions update.} Elsevier, 2022, pp. 217--244.

  \bibitem{Jamshid}
{O. Moztarzadeh, M. Jamshidi, S. Sargolzaei, A. Jamshidi, N. Baghalipour, M. M. Moghani, and L. Hauer. "Metaverse and healthcare: Machine learning-enabled digital twins of cancer," \emph{Bioengineering,} vol. 10, no. 4, Apr. 2023.}
  
  
  \bibitem{arxive} 
		M. Letafati, H. Behroozi, B. H. Khalaj, and E. A. Jorswieck, ``On learning-assisted content-based secure image transmission for delay-aware systems with randomly-distributed eavesdroppers,'' \emph{IEEE Trans.  Commun.},  vol. 70, no. 2, pp. 1125--1139, Feb. 2022.
		%doi: 10.1109/TCOMM.2021.3128423, Nov. 2021. 
		
		%\textcolor{gray}
		%{\bibitem{twc}
		%	M. 	Letafati, A. Kuhestani, H. Behroozi and D. W. K. Ng, ``Jamming-resilient frequency hopping-aided secure communication for Internet-of-Things in the presence of an untrusted relay,'' \emph{IEEE Trans. on Wireless Commun.}, vol. 19, no. 10, pp. 6771--6785, Oct. 2020.}
		
	\bibitem{ICC}
				M. Letafati, H. Behroozi, B. H. Khalaj and E. A. Jorswieck, ``Content-based medical image transmission against randomly-distributed passive eavesdroppers,'' \emph{2021 IEEE International Conference on Communications Workshops (ICC Workshops),} Montreal, QC, Canada, 2021, pp. 1-7. 
		 
		
		\bibitem{WSKG-GC} 
		M. Letafati, H. Behroozi, B. H. Khalaj, and E. A. Jorswieck,   ``Deep learning for hardware-impaired wireless
		secret key generation with man-in-the-middle attacks,'' \emph{2021 IEEE Global Communications Conference (GLOBECOM),} Madrid, Spain, Dec. 2021, pp. 1--6.  	
		
		\bibitem{vtc2022}
		M. Letafati, H. Behroozi, B. H. Khalaj, and E. A. Jorswieck,   ``Wireless-powered cooperative key generation for e-health: A reservoir learning approach,'' \emph{2022 IEEE 95th Vehicular Technology Conference (VTC-Spring),} Helsinki, Finland, Jun. 2022, pp. 1--7. 	


\bibitem{battle_privacy} 
		B. Falchuk, S. Loeb, and R. Neff, ``The social metaverse: Battle for
		privacy,'' \textit{IEEE Technology and Society Magazine}, vol. 37, no. 2, pp.
		52–61, 2018.	

\bibitem{medical_deepfake}
		S.  Solaiyappan and Y. Wen, ``Machine learning based medical image deepfake detection: A comparative study,'' \emph{arXiv:2109.12800v2}, Apr. 2022.  
		

  \bibitem{model_inversion} 
		J. Geiping, H. Bauermeister, H. Dröge, M. Moeller, ``Inverting Gradients - How easy is it to break privacy in federated learning?,'' \emph{Advances in Neural Information Processing Systems (NeurIPS)}, Dec. 2020. 
		%		M. Fredrikson, S. Jha, and T. Ristenpart, ``Model inversion attacks that
		%		exploit confidence information and basic countermeasures,'' in \emph{Proc.
		%		ACM CCS}, New York, NY, USA, 2015, pp. 1322--1333. 
		%		

  \bibitem{backdoor} 
		H. Wang, et al., ``Attack of the tails: Yes, You really can backdoor federated learning,'' 	  \emph{Advances in Neural Information Processing Systems (NEURIPS 2020)}, Dec. 2020,   pp. 16070--16084. 


  %\bibitem{BC_FL_coop}
	%	L. Jiang, H. Zheng, H. Tian, S. Xie, and Y. Zhang, ``Cooperative federated
		%learning and model update verification in blockchain empowered
		%digital twin edge networks,'' \textit{IEEE Internet of Things Journal,} vol. 9, no. 13, Jul. 2022. 
		
		\bibitem{BC_FL_core}
		J. Kang, D. Ye, J. Nie, J. Xiao, X. Deng, S. Wang,
		Z. Xiong, R. Yu, and D. Niyato ``Blockchain-based federated learning for industrial 
		metaverses: Incentive scheme with optimal AoI,''  
  \emph{2022 IEEE International Conference on Blockchain (Blockchain)}, Espoo, Finland, Aug. 2022, pp. 71-78. 
   %\textit{arXiv preprint arXiv:2206.07384}, Jun. 2022.
   

	\bibitem{Vivek3}
	V. Nair, W. Guo, J. Mattern, R. Wang, J. F. O'Brien, L. Rosenberg, and D. Song, ``Unique identification of 50000+ virtual reality users from head \& hand motion data,'' 	
	\textit{arXiv:2302.08927}, Feb. 2023.
  
		\bibitem{Vivek1}	
	V. Nair, G. M. Garrido, and D. Song, ``Exploring the unprecedented privacy risks of the metaverse,''   \textit{arXiv:2207.13176v1}, Jul. 2022. 
	
	\bibitem{Vivek2}
	V. Nair, G. M. Garrido, and D. Song, ``Going incognito in the metaverse,'' \textit{arXiv:2208.05604v1}, Aug. 2022. 	

	
  \bibitem{Dopamine} 
  M. Malekzadeh, B. Hasircioglu, N.  Mital, K.  Katarya, M. E.   Ozfatura,   and D.  Gündüz, ``Dopamine: Differentially private federated learning on medical data,'' \emph{The Second AAAI Workshop on Privacy-Preserving Artificial Intelligence (PPAI-21)}, Feb. 2021, pp. 1--9.    
		
		%\bibitem{Washington_post}
		%T. Hunter, ``\emph{Surveillance will follow us into ‘the metaverse,’ and our bodies could be its new data source.''} Washington Post, Jan. 2022, https://www.washingtonpost.com/technology/2022/01/13/privacy-vr-metaverse/, [Accessed Aug. 29, 2022].
		
		
		\bibitem{synthetic}
		S. 	Castellanos, (2021, July 23). ``\textit{Fake It to Make It: 	Companies Beef Up AI Models With Synthetic Data.}'' WSJ:  \href{https://www.wsj.com/articles/fake-it-to-make-it-companies-beef-up-ai-models-with-synthetic-data-11627032601}{https://www.wsj.com/articles/fake-it-to-make-it-companies-beef-up-ai-models-with-synthetic-data-11627032601},  [Accessed Jun. 5, 2023].  
		
		
		%\bibitem{gan_people}
		%S. Nightingale, J. F. Hany, ``AI-synthesized faces are indistinguishable from real faces and more trustworthy,'' in \emph{Proceedings of the National Academy of Sciences,} Feb. 2022.	


		%\bibitem{accenture}  
		%P. Daugherty, M. Carrel-Billiard, and M. Biltz, \emph{``Meet me in the metaverse:‌ The continuum of technology and experience, reshaping business.''} 
		%https://www.accenture.com/us-en/insights/technology/technology-trends-2022, [Accessed Aug. 28, 2022].  
	
        \bibitem{SecAgg} 
		J. So, C. He, C. -S. Yang, S. Li, Y. Qian, and S. Avestimehr,   ``LightSecAgg: A lightweight and versatile design for secure aggregation in federated learning,''  \emph{Proceedings of Machine Learning and Systems (MLSys 2022),}  \emph{arXiv:2109.14236v3}, Feb. 2022. [Online]. Available: https://arxiv.org/abs/2109.14236.

  
\bibitem{ai.googleblog}
f.  Hartmann and P. Kairouz, (2023, March 2). ``\textit{Distributed differential privacy for federated learning.}'' Google Research: 
\url{https://ai.googleblog.com/2023/03/distributed-differential-privacy-for.html},  [Accessed Jun. 5, 2023]. 


\bibitem{WSKG-letter} 
		M. Letafati, H. Behroozi, B. H. Khalaj, and E. A. Jorswieck, ``Hardware-impaired
		PHY secret key generation with man-in-the-middle adversaries,''
		\emph{IEEE Wireless Commun. Lett.,} vol. 11, no. 4, pp. 856--860, Apr. 2022.


\bibitem{FL_Quek}
		K. Wei, J. Li, M. Ding, C. Ma, H. H. Yang, F. Farokhi, S. Jin,
		T. Q. Quek, and H. V. Poor, “Federated learning with differential
		privacy: Algorithms and performance analysis,” \textit{IEEE Transactions on Information Forensics and Security}, vol. 15, pp. 3454--3469, 2020.

 \bibitem{AML_tomg}
		J. Geiping, L. H Fowl, G. Somepalli, M.  Goldblum,  M. Moeller, and T. Goldstein,	``What doesn't kill you makes you robust(er): How to adversarially train against data poisoning,''    \textit{International Conference on Learning Representations (ICLR),} Apr.  2022. 
		
		%\bibitem{sover}
		%K. Schmidt, G. M. Garrido, A. Mühle, and C.  Meinel, ``Mitigating sovereign data exchange challenges:
		%A mapping to apply privacy- and
		%authenticity-enhancing technologies,''  
		%\textit{arXiv:2207.01513v1}, Jun. 2022. 


  \bibitem{Gaussian}
C. Dwork and A. Roth, ``The algorithmic foundations of differential privacy,'' 
\emph{Found. Trends Theor. Comput. Sci.,} vol. 9, nos. 3--4, pp. 211--407, 2013.
  


  \bibitem{accenture}  
		P. Daugherty, M. Carrel-Billiard, and M. Biltz, \emph{``Meet me in the metaverse:‌ The continuum of technology and experience, reshaping business.''} 
		\url{https://www.accenture.com/us-en/insights/technology/technology-trends-2022}, [Accessed Jun. 8, 2023].  
		
		%\bibitem{DBA}
		%C. Xie, K. Huang, P.- Y. Chen, and B. Li,  ``DBA: Distributed backdoor attacks against federated learning,'' \emph{International Conference on Learning Representations (ICLR),}  Apr. 2020, Ethiopia. 
	
		
		
		%\bibitem{AML_social_media}
		%V. Cherepanova, M. Goldblum, H. Foley, S. Duan, J. P. Dickerson, Ga. Taylor,  and T. Goldstein, ``LowKey: Leveraging adversarial attacks to protect social media users from facial recognition,'' \textit{International Conference on Learning Representations (ICLR),} May 2021. 
		
		
		%\bibitem{personal_space} 
		%L. E. Buck and B. Bodenheimer, ``Privacy and personal space: Addressing interactions and interaction data as a privacy concern,'' \emph{2021 IEEE Conference on Virtual Reality and 3D User Interfaces Abstracts and Workshops (VRW)}, Lisbon. Portugal,  Apr. 2021,  pp. 399--400. 


\bibitem{AE_paper}
S. A. A. Kalkhoran, M. Letafati, E. Erdemir, B. H. Khalaj, H. Behroozi, and D. Gündüz, ``Secure deep-JSCC against multiple eavesdroppers,''  
 \emph{arXiv:2308.02892}, Aug. 2023. [Online]. Available: https://arxiv.org/abs/2308.02892. 


\bibitem{Centralized_SVM_Wisconsin}
J. Liang, Z. Qin, J. Ni, X. Lin, and X. Shen, ``Practical and secure SVM classification for cloud-based remote clinical decision services,'' \emph{IEEE Transactions on Computers,} vol. 70, no. 10, pp. 1612--1625, Oct. 2021. 

		
		%\bibitem{human}
		%L. Buck  and R. McDonnell, ``Security and privacy in the metaverse: The threat of the digital human,'' \emph{Proceedings of the 1st Workshop on Novel Challenges of Safety, Security and 	Privacy in Extended Reality (CHI EA'22),}  New Orleans, LA, USA, May 2022, pp. 1--4.   
		
		
		%\bibitem{ericsson-IoSense}
		%``\emph{10 Hot Consumer Trends 2030: The Internet of Senses.}'' https://www.ericsson.com/en/reports-and-papers/consumerlab/reports/10-hot-consumer-trends-2030. [Accessed Sep. 2, 2022].

\bibitem{dataset}
{O. Moztarzadeh, M. Jamshidi, S. Sargolzaei, F. Keikhaee, A. Jamshidi, S. Shadroo, and L. Hauer, "Metaverse and medical diagnosis: A blockchain-based digital twinning approach based on MobileNetV2 algorithm for cervical vertebral maturation" \emph{Diagnostics}, vol. 13, no. 8, Apr. 2023.}

\end{thebibliography}
\end{document}